\documentclass[journal=aelccp,manuscript=letter]{achemso}

\usepackage{graphicx}
\usepackage{bm}
\usepackage{multirow}
\usepackage[colorlinks=true,linkcolor=blue,urlcolor=blue, citecolor=blue]{hyperref}
\usepackage{array}
\usepackage{hyperref}
\usepackage{amsmath,amssymb}
\usepackage{chemformula} 
\usepackage{booktabs}

\graphicspath{{./Figures/}}

\author{Haibo Xue}
    \affiliation{Materials Simulation \& Modelling, Department of Applied Physics, Eindhoven University of Technology, P.O. Box 513, 5600MB Eindhoven, the Netherlands.}
    \alsoaffiliation{Center for Computational Energy Research, Department of Applied Physics, Eindhoven University of Technology, P.O. Box 513, 5600MB Eindhoven, the Netherlands.}

\author{Zehua Chen}
    \affiliation{Materials Simulation \& Modelling, Department of Applied Physics, Eindhoven University of Technology, P.O. Box 513, 5600MB Eindhoven, the Netherlands.}
    \alsoaffiliation{Center for Computational Energy Research, Department of Applied Physics, Eindhoven University of Technology, P.O. Box 513, 5600MB Eindhoven, the Netherlands.}
    
\author{Shuxia Tao}
    \affiliation{Materials Simulation \& Modelling, Department of Applied Physics, Eindhoven University of Technology, P.O. Box 513, 5600MB Eindhoven, the Netherlands.}
    \alsoaffiliation{Center for Computational Energy Research, Department of Applied Physics, Eindhoven University of Technology, P.O. Box 513, 5600MB Eindhoven, the Netherlands.}
    \email{s.x.tao@tue.nl}
    
\author{Geert Brocks}
    \affiliation{Materials Simulation \& Modelling, Department of Applied Physics, Eindhoven University of Technology, P.O. Box 513, 5600MB Eindhoven, the Netherlands.}
    \alsoaffiliation{Center for Computational Energy Research, Department of Applied Physics, Eindhoven University of Technology, P.O. Box 513, 5600MB Eindhoven, the Netherlands.}
    \alsoaffiliation{Computational Materials Science, Faculty of Science and Technology and MESA+ Institute for Nanotechnology, University of Twente, P.O. Box 217, 7500AE Enschede, the Netherlands.}
    \email{g.h.l.a.brocks@utwente.nl}

\title{Defects in Halide Perovskites:\\ Does It Help to Switch from 3D to 2D?}


\begin{document}

\begin{tocentry}

\includegraphics{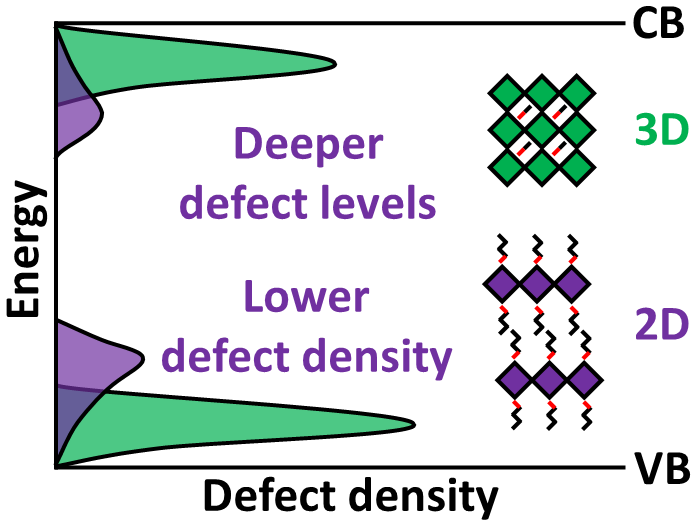}

\end{tocentry}

\newpage

\begin{abstract}
Ruddlesden-Popper hybrid iodide 2D perovskites are put forward as stable alternatives to their 3D counterparts. Using first-principles calculations, we demonstrate that equilibrium concentrations of point defects in the 2D perovskites PEA$_2$PbI$_4$,  BA$_2$PbI$_4$, and PEA$_2$SnI$_4$ (PEA: phenethyl ammonium, BA: butylammonium), are much lower than in comparable 3D perovskites. Bonding disruptions by defects are more detrimental in 2D than in 3D networks, making defect formation energetically more costly. The stability of 2D Sn iodide perovskites can be further enhanced by alloying with Pb. Should, however, point defects emerge in sizable concentrations as a result of nonequilibrium growth conditions, for instance, then those defects hamper the optoelectronic performance of the 2D perovskites, as they introduce deep traps. We suggest that trap levels are responsible for the broad sub-bandgap emission in 2D perovskites observed in experiments.
   
\end{abstract}

\clearpage

\section{Manuscript}

Hybrid organometal halide perovskites are materializing as candidate semiconductors for new generations of optoelectronic devices such as solar cells and light-emitting diodes \cite{Jena2019, Saparov2016}. Application of these materials, however, is severely hampered by their lack of long-term stability \cite{Kim2016, Zhouyuanyuan2019, Park2019, He2020}. One of the first frequently studied compounds, MAPbI$_3$, has favorable optical and charge transport properties \cite{Kojima2009, Quarti2016, Haruyama2016, Holzhey2019}, but the MA$^+$ (methylammonium) ion is chemically not sufficiently stable, and suffers from degradation reactions \cite{Juarez-Perez2016, Conings2015}. Replacing MA$^+$ by larger and more stable cations, such as FA$^+$ (formamidinium) \cite{Eperon2014} or GA$^+$ (guanidinium) \cite{Giorgi2015}, suffers from the perovskite structure becoming unstable, leading to a tendency to convert to different crystal structures that are much less optically active \cite{Stoumpos2013, Han2016, Jodlowski2016}. This tendency can be suppressed to a certain extent by mixing in smaller inorganic cations, such as Cs$^+$ \cite{Liu2017, Yi2016}, but the fundamental issue remains that a stable 3D perovskite lattice requires the sizes of the constituting ions to be of a certain proportion, as expressed by the Goldschmidt tolerance factor \cite{Goldschmidt1926, Saparov2016}, and the scale is set by the 3D network of metal halide octahedra in the perovskite.

In recent years, organometal halide perovskites with a Ruddlesden-Popper structure have emerged as alternative materials \cite{Stoumpos2016}. In these perovskites the metal halide octahedra form a planar 2D network, and these 2D layers are separated by layers of organic cations, where the interlayer interaction is typically Vanderwaals \cite{Blancon2020}. Using organic ions with a quasi-linear structure, such as PEA (phenethylammonium)\cite{Calabrese1991} or BA (butylammonium) \cite{Mitzi1994}, the in-plane tolerance factor for a stable crystal structure is easily obeyed, whereas the out-of-plane size of the organic ion becomes relatively unimportant. Although the stability of such 2D perovskites is markedly improved, as compared to their 3D counterparts, presently photoelectric devices based upon 2D perovskites fail to reach the high efficiencies obtained with 3D perovskites \cite{Blancon2020}. In terms of this, defects can play an important role, whereas their concentrations in 2D perovskites and resulting impacts on electronic properties are not yet clear \cite{Blancon2020}.

In this paper we explore the defect chemistry and physics of prominent 2D organometal iodide perovskites, PEA$_2$PbI$_4$,  BA$_2$PbI$_4$, and PEA$_2$SnI$_4$, and the alloy PEA$_2$Sn$_{0.5}$Pb$_{0.5}$I$_4$, using first-principles density functional theory (DFT) calculations. The ease with which point defects can be created in a material is an indication for its stability. We therefore focus on the defect formation energy (DFE) as it can be calculated assuming thermodynamic equilibrium conditions. We use the same formalism as applied in our previous work on 3D perovskites \cite{Xue2022perovskites}. A summary of the theory is given in the Supporting Information (SI), Sec. I. The equilibrium chemical potentials of the different elements are determined by considering the phase diagram of the 2D perovskite, see the SI, Fig. S1. The defect formation energies are calculated using iodine-medium conditions, which are the conditions most typically used.

Even if defects do not occur in large quantities under thermodynamic equilibrium conditions, they may appear more prominently under nonequilibrium growth conditions, or under operating conditions \cite{Xue2022_compund_defects}. If so, they can seriously affect the electronic properties of the material, as defect states with energy levels inside the semiconductor band gap can act as traps for charge carriers, and as recombination centers for radiationless decay. We explore these energy levels, called charge-state transition levels (CSTLs), associated with the most likely point defects in the 2D materials listed above.

\begin{figure} [t]
    \includegraphics[width=0.9\textwidth]{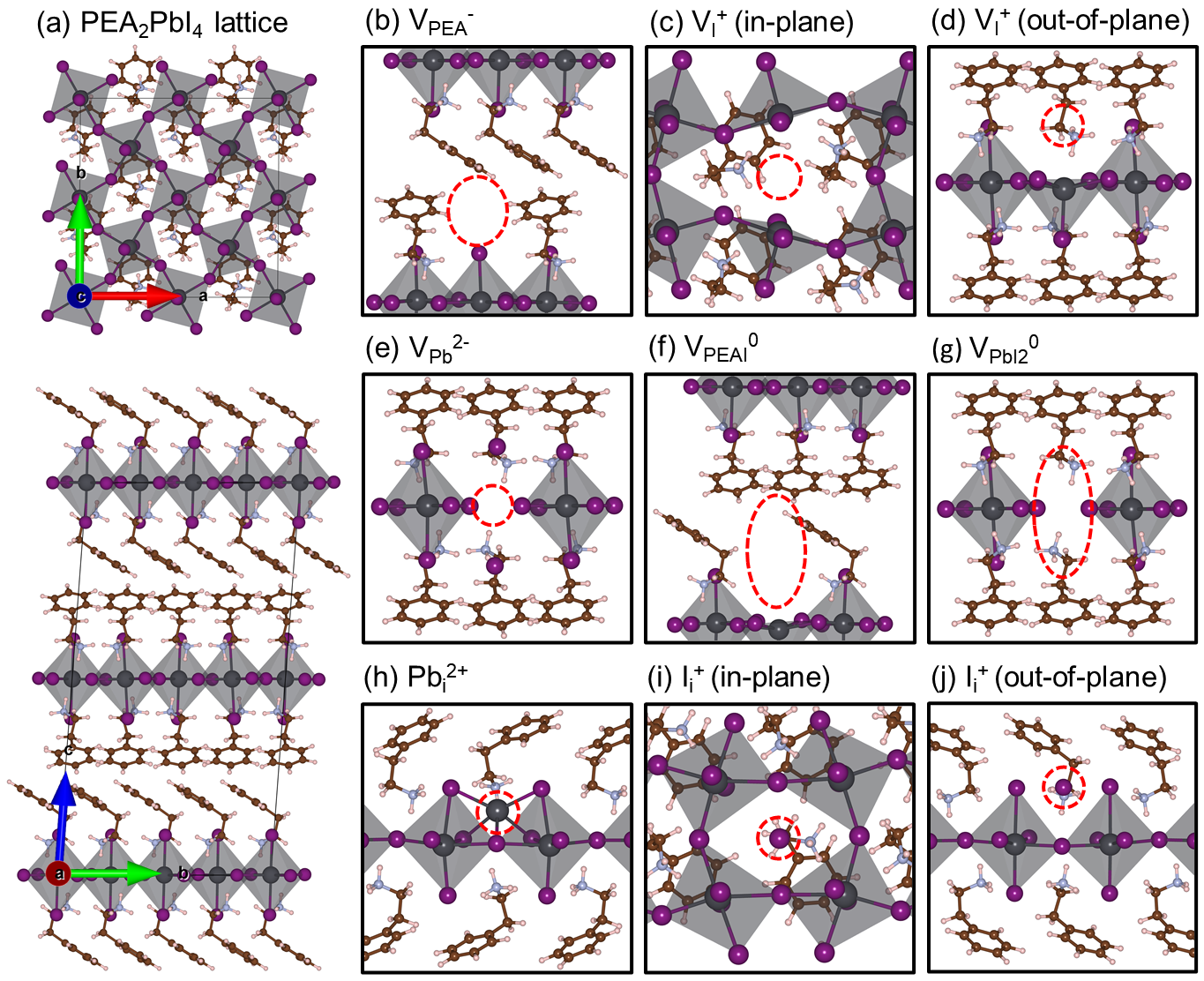}
    \caption{(a) Top and side views of a $2\times2\times2$ PEA$_2$PbI$_4$ supercell; optimized structures of vacancies (b-g) and interstitials (h-j) in their most stable charge states in PEA$_2$PbI$_4$. The positions of the defects are marked in red. The labels (in-plane) and (out-of-plane) refer to positions of iodine vacancies and interstitials either within an PbI$_2$ plane or above/below it.}
    \label{fig: structures}
\end{figure}

DFT calculations are performed on $2 \times 2 \times 1$ supercells of 2D perovskites, with the Vienna \textit{Ab Initio} Simulation Package (VASP) \cite{Kresse1993, Kresse1996, Kresse1996a}, employing the SCAN + rVV10 functional \cite{Peng2016} for electronic calculations and geometry optimization. The SCAN+rVV10 functional is used aiming at obtaining accurate defective structures and total energies, and therefore the DFEs, which are the main focus of this work. Whereas the DFT band gap error may result in incorrect band edges, the calculated CSTLs are suggested to be correct in a relative sense, as discussed in our previous work Ref. \citenum{Xue2021functionals}. Detailed computational settings and structures are discissed in the SI, Sec. I. We start with point defects in the most popular 2D perovskite, PEA$_2$PbI$_4$,  i.e, the PEA vacancy $\mathrm{V_{PEA}}$, the Pb vacancy $\mathrm{V_{Pb}}$, and the iodine vacancy $\mathrm{V_{I}}$, and the interstitials $\mathrm{Pb_i}$ and $\mathrm{I_i}$. The PEA interstitial is omitted because the structure is too dense for additionally accommodating such an extra large-size organic cation. In addition to these simple point defects, we also study the compound vacancies $\mathrm{V_{PEAI}}$ and $\mathrm{V_{PbI_2}}$, representing missing units of the precursors PEAI and PbI$_2$. The layered nature of the 2D perovskite [\autoref{fig: structures}(a)], implies that iodine vacancies and interstitials in the central PbI planes and those outside these planes can behave differently, and both configurations are studied. Optimized structures of all defects in their most stable charge states are shown in \autoref{fig: structures}(b-j).

\begin{figure}[t]
    \includegraphics[width=1\textwidth]{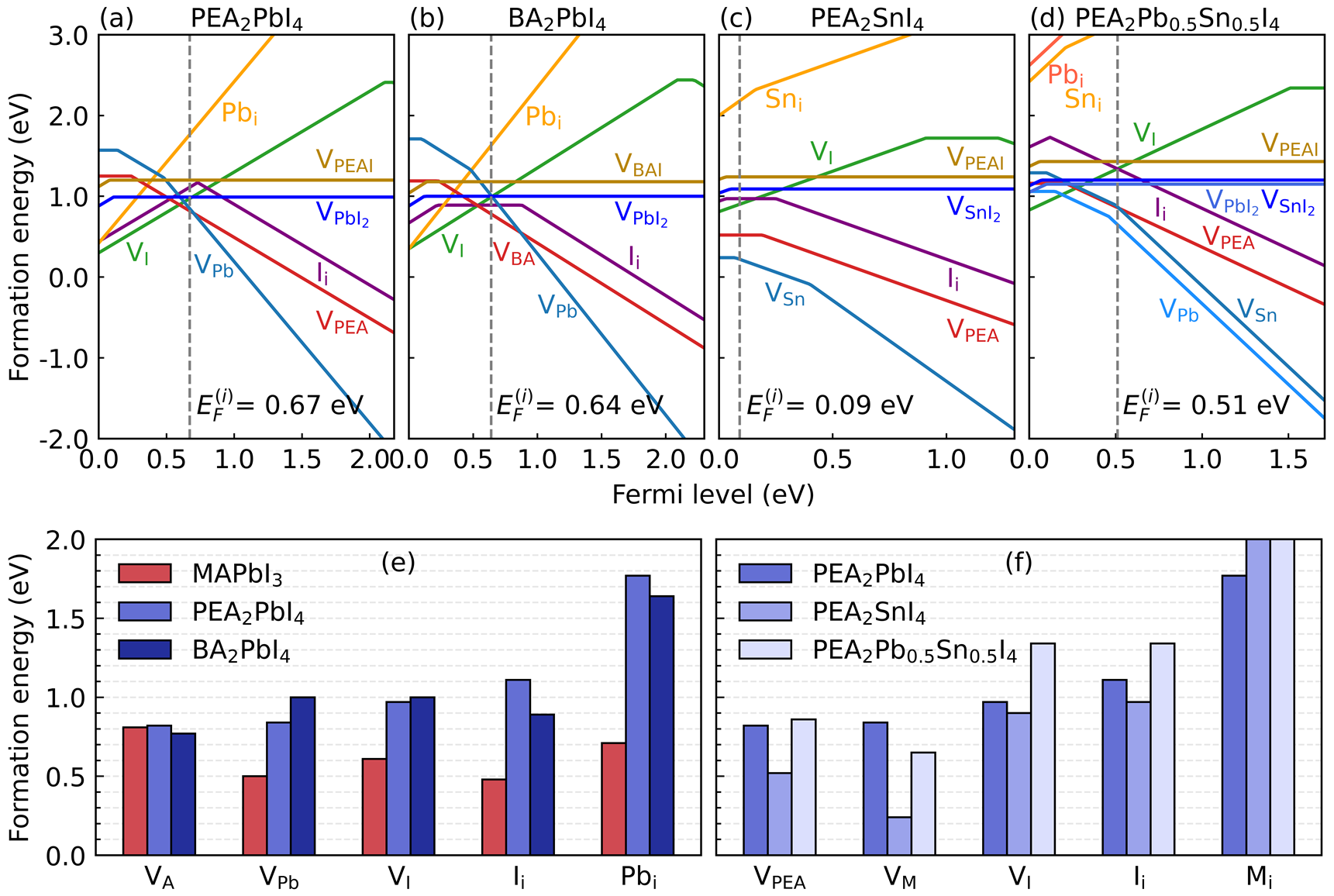}
    \caption{(a-d) Defect formation energies in PEA$_2$PbI$_4$, BA$_2$PbI$_4$, PEA$_2$SnI$_4$, and PEA$_2$Pb$_{0.5}$Sn$_{0.5}$I$_4$ as function of the position of the Fermi level; the intrinsic Fermi level is indicated by the black dashed line. Comparison of DFEs at the intrinsic Fermi level to MAPbI$_3$ (e), and among 2D perovskites with different metal cations (f); the DFEs of $\mathrm{Sn_i}$ are 2.2 eV and 3.1 eV, for PEA$_2$SnI$_4$ and PEA$_2$Pb$_{0.5}$Sn$_{0.5}$I$_4$, respectively.}
    \label{fig: DFEs}
\end{figure}

The calculated DFEs of PEA$_2$PbI$_4$ are shown in \autoref{fig: DFEs}(a). The intrinsic Fermi level ($E_F^{(i)}= 0.67$ eV with respect to the valence band maximum, VBM) is obtained from the charge neutrality condition, see SI, Sec. I.3. At this condition, the vacancies $\mathrm{V_{PEA}}^-$ and $\mathrm{V_{Pb}}^{2-}$ are easiest to form, and thereby are the most dominant defects, with formation energies of 0.82 eV and 0.84 eV, respectively. This leads to equilibrium concentrations at room temperature of 5.73 $\times 10^7$ cm$^{-3}$ and 1.09 $\times 10^7$ cm$^{-3}$, respectively. Other vacancies, $\mathrm{V_I}^+$, and the compound vacancies $\mathrm{V_{PEAI}}^0$ and $\mathrm{V_{PbI_2}}^0$, as well as all interstitial species, have formation energies $\gtrsim 1$ eV, and are thus unimportant under equilibrium conditions (with the intrinsic Fermi level). A full list of formation energies and concentrations at room temperature of defects is given in \autoref{table: DFE and concentration of charged defects}.

\begin{table}[h] 
    \caption{ Formation energies $\Delta H_f$ (eV) and concentrations $c$ (cm$^{-3}$) at $T$ = 300 K of charged defects in 3D and 2D halide perovskites calculated at the intrinsic Fermi level, and iodine-medium conditions, using the SCAN+rVV10 functional. Formation energies and concentrations of the dominant defects in each perovskite are underlined. For those defects which are stable at a different charge state at the intrinsic Fermi level, the corresponding stable charge state is specified in the brackets.}
    \label{table: DFE and concentration of charged defects}
    \renewcommand{\arraystretch}{1.2} 
        \begin{tabular}{lllllll}
            \toprule
                          & $\mathrm{V_{A}}^-$ & $\mathrm{V_{M}}^{2-}$ & $\mathrm{V_{I}}^+$ & $\mathrm{I_i}^-$  & $\mathrm{M_i}^{2+}$\\
            \midrule
            \multicolumn{6}{l}{Defect formation energy $\Delta H_f$ (eV)}\\
            MAPbI$_3$ & 0.81 & \underline{0.50} & 0.61 & \underline{0.48} & 0.71  \\
            PEA$_2$PbI$_4$ & \underline{0.82} & \underline{0.84} & 0.97 & 1.11 ($\mathrm{I_i}^+$) & 1.77\\
            BA$_2$PbI$_4$ & \underline{0.77} & 1.00 & 1.00 & 0.89 ($\mathrm{I_i}^0$) & 1.64\\
                MASnI$_3$ \textsuperscript{\emph{a}} & 0.35 & \underline{$-$0.21} & 0.92 & 0.16 & 0.81\\
            PEA$_2$SnI$_4$ & 0.52 ($\mathrm{V_{PEA}}^0$) & \underline{0.24} ($\mathrm{V_{Sn}}^{-}$) & 0.90 & 0.97 ($\mathrm{I_i}^0$) & 1.28 ($\mathrm{Sn_i}^{+}$)\\
            PEA$_2$Pb$_{0.5}$Sn$_{0.5}$I$_4$ & 0.86 & \underline{0.65} & 1.34 & 1.34 & 3.15 ($\mathrm{Sn_i}^{+}$)\\
            \multicolumn{6}{l}{Defect concentration $c$ (cm$^{-3}$)}\\
            MAPbI$_3$ & 8.68$\times 10^{7}$ & \underline{1.44$ \times$ 10$^{13}$} & 6.10$\times 10^{11}$ & \underline{1.37 $\times$ 10$^{14}$} & 6.36$\times 10^{10}$ \\
            PEA$_2$PbI$_4$ & \underline{5.37 $\times$ 10$^{7}$} & \underline{1.09 $\times$ 10$^{7}$} & 3.27$\times 10^{5}$ & 3.79$\times 10^{2}$ & 4.02$\times 10^{-8}$ \\
            BA$_2$PbI$_4$ & \underline{4.41 $\times$ 10$^{8}$} & 3.68$\times 10^{4}$ & 1.38$\times 10^{5}$ & 1.80$\times 10^{6}$ & 5.86$\times 10^{-6}$ \\
            MASnI$_3$ \textsuperscript{\emph{a}} & $5.96\times10^{15}$ & \underline{1.30$\times$ 10$^{25}$} & 4.64$\times 10^{6}$ & 2.12$\times 10^{19}$ & 3.35$\times 10^{0}$ \\
            PEA$_2$SnI$_4$ & 3.13$\times 10^{12}$ & \underline{3.08 $\times$ 10$^{17}$} & 4.21$\times 10^{6}$ & 3.09$\times 10^{5}$ & 4.88$\times 10^{-15}$ \\
            PEA$_2$Pb$_{0.5}$Sn$_{0.5}$I$_4$ & 1.16$\times 10^{7}$ & \underline{2.30$\times$ 10$^{10}$} & 2.18$\times 10^{-1}$ & 1.59$\times 10^{-1}$ & 2.95$\times 10^{-31}$ \\
            \bottomrule
        \end{tabular}
        \textsuperscript{\emph{a}}To facilitate comparison to the other compounds, formation energies and defect concentrations in MASnI$_3$ are calculated under iodine-medium conditions, which are however unrealistic experimentally for this compound  \cite{Xue2022perovskites}.
\end{table}

These findings are in stark contrast with results obtained for 3D perovskites, calculated using the same computational settings  \cite{Xue2022perovskites}.  In the archetype 3D perovskite MAPbI$_3$, the formation energies of several point defects are $\lesssim 0.5$ eV, which leads to equilibrium defect concentrations that are seven orders of magnitude higher at room temperature than in PEA$_2$PbI$_4$ \cite{Xue2022perovskites}, \autoref{fig: DFEs}(e). Moreover, interstitials generally play a significant role in the defect chemistry of 3D perovskites\cite{Xue2022perovskites}. In MAPbI$_3$, besides the vacancy $\mathrm{V_{Pb}}^{2-}$, the dominant point defects are the interstitials $\mathrm{MA_i}^+$ and $\mathrm{I_i}^-$.  By comparison, interstitials in PEA$_2$PbI$_4$ are unimportant relative to vacancies.

The chemical bonding patterns of such interstitials in 2D perovskites are also quite different from those in 3D. Iodine interstitials in 3D perovskites are inserted between two Pb atoms in the lattice, next to an already present iodine ion, forming a Pb-I$_2$-Pb unit with two equivalent Pb-I-Pb bridge bonds.  \cite{Meggiolaro2018iodine,Xue2022perovskites}. In contrast, iodine interstitials in PEA$_2$PbI$_4$ prefer to stay in-between two lattice iodines, forming a $\mathrm{I_3}$ trimer structure, \autoref{fig: structures}(i),(j). This difference in bonding pattern is also reflected in the most stable charge state. Whereas the iodine interstitial in 3D perovskites is negatively charged, in PEA$_2$PbI$_4$ it is positively charged. Alternatively, one might call the trimer structure with the lattice iodines a negatively charged $\mathrm{I_3}^-$ unit.

The more prominent defects in PEA$_2$PbI$_4$, the vacancies $\mathrm{V_{Pb}}^{2-}$ and $\mathrm{V_{PEA}}^-$, \autoref{fig: structures}(b),(e), and also the less prominent vacancies $\mathrm{V_{I}}^{+}$, $\mathrm{V_{PEAI}}^0$ and $\mathrm{V_{PbI_2}}^0$, \autoref{fig: structures}(c),(d),(f),(g), have bonding patterns that are qualitatively similar to those in 3D perovskites, and stable charge states that are the same. The DFEs of these defects in PEA$_2$PbI$_4$ are significantly higher, though. Whereas in 3D perovskites the presence of a vacancy can be accommodated to some extent by rearranging the lattice around the vacancy, in a 2D perovskite such a rearrangement is more difficult. In addition, creating charged defects in 2D perovskites might be more difficult in general, because of the smaller dielectric constants of these materials \cite{Leslie1985,Makov1995,Traore2018}. 

The analysis of results obtained for PEA$_2$PbI$_4$ also holds qualitatively for other 2D perovskites, such as BA$_2$PbI$_4$, whose DFEs are shown in \autoref{fig: DFEs}(b). Comparing to PEA$_2$PbI$_4$, it is observed that the formation energies of defects in BA$_2$PbI$_4$ are 0.05-0.15 eV different from the corresponding ones in PEA$_2$PbI$_4$, \autoref{fig: DFEs}(e). The vacancy $\mathrm{V_{BA}}^-$ has a somewhat smaller DFE than the vacancy $\mathrm{V_{PEA}}^-$, 0.77 eV versus 0.82 eV, which likely reflects the fact that removing the smaller BA$^+$ leaves a smaller hole in the lattice. As for the other defects, vacancies in BA$_2$PbI$_4$ tend to be somewhat destabilized, as compared to those in PEA$_2$PbI$_4$, whereas interstitials are somewhat stabilized. None of these changes affect the qualitative comparison to 3D perovskites as discussed above, however.

Stabilization of compounds in 2D structures offers an interesting perspective for Sn-based perovskites, where for instance the 3D MASnI$_3$ perovskite is quite unstable. That is reflected by the ease with which $\mathrm{V_{Sn}}^{2-}$ vacancies are generated spontaneously, making this compound an intrinsically doped degenerate p-type semiconductor \cite{Xue2022perovskites}. In the literature this is related to the fact that Sn$^{2+}$ can be easiliy oxidized to Sn$^{4+}$ \cite{Noel2014, Tao2019, Dalpian2017, Leijtens2017}. In fact, calculations indicate that MASnI$_3$ is thermodynamically stable only under rather extreme iodine-poor conditions  \cite{Xue2022perovskites}.

\autoref{fig: DFEs}(c) shows the formation energies of defects in PEA$_2$SnI$_4$, calculated under milder, iodine-medium, conditions. All DFEs are positive, indicating that the material is stable against the spontaneous formation of defects, which is in contrast with MASnI$_3$, see SI Table 1. The vacancy $\mathrm{V_{Sn}}$ in PEA$_2$SnI$_4$ is the defect with lowest DFE, but the latter is considerably higher than its DFE in MASnI$_3$, signaling an increased stability of the material. The vacancy is negatively charged, $\mathrm{V_{Sn}}^-$, and, as is common in Sn-based halide perovskites, there is no appreciable concentration of positively charged defects to maintain charge neutrality  \cite{Meggiolaro2020PbVsSn,Xue2022perovskites}. The latter has to be ensured by holes in the valence band, which leads to an intrinsic Fermi level that is only 0.09 eV above the VBM. It results in PEA$_2$SnI$_4$ being an intrinsic $p$-type semiconductor, albeit not a degenerate one, as is the case for MASnI$_3$.   

The DFE of the main defect, $\mathrm{V_{Sn}}^-$, in PEA$_2$SnI$_4$ is 0.26 eV, which gives an equilibrium concentation (at room temperature) of $3.08\times 10^{17}$ cm$^{-3}$. The vacancy $\mathrm{V_{PEA}}^0$ has a DFE of 0.52 eV, and an equilibrium concentration of $3.13 \times 10^{12}$ cm$^{-3}$, whereas other vacancies and interstitials have a DFE in excess of 0.9 eV, so they do not play a significant role. Compared to PEA$_2$PbI$_4$, note that the most stable charge states of the prominent defects are $+1e$ higher ($\mathrm{V_{Pb}}^{2-}$ and $\mathrm{V_{PEA}}^-$). This stems from the low lying intrinsic Fermi level in PEA$_2$SnI$_4$, as compared to that in PEA$_2$PbI$_4$, consistent with an increased p-type doping, comparing \autoref{fig: DFEs}(a) and (c).

One can observe that the DFEs of defects in PEA$_2$SnI$_4$ tend to be significantly smaller than the corresponding ones in PEA$_2$PbI$_4$, see \autoref{fig: DFEs}(f). The intrinsic defect concentrations in the former are therefore much higher, which is a clear sign of a decrease in stability. In particular, 2D Sn-based perovskites inherit the problem of a large amount of Sn vacancies from the 3D Sn-based perovskites. To suppress the formation of Sn vacancies, one solution is to mix Sn with Pb \cite{Meggiolaro2020PbVsSn, Leijtens2017}. We investigate its effect using the perovskite $\mathrm{PEA_2Pb_{0.5}Sn_{0.5}I_4}$. As an example, we study a highly ordered structure of $\mathrm{PEA_2Pb_{0.5}Sn_{0.5}I_4}$, whose construction is discussed in Figure S2 of the Supporting Information. The calculated DFEs are shown in \autoref{fig: DFEs}(d). It is evident that it is more difficult to form defects in $\mathrm{PEA_2Pb_{0.5}Sn_{0.5}I_4}$ as compared the pure Sn-based perovskite PEA$_2$SnI$_4$. In addition, the intrinsic Fermi level in $\mathrm{PEA_2Pb_{0.5}Sn_{0.5}I_4}$ is calculated at 0.51 eV, making it a mildly doped intrinsic p-type semiconductor, which, compared to PEA$_2$SnI$_4$, is more in line with the other 2D perovskites, see \autoref{fig: DFEs}(a)-(d). 

The defect in $\mathrm{PEA_2Pb_{0.5}Sn_{0.5}I_4}$ that is easiest to form is the vacancy $\mathrm{V_{Pb}}^{2-}$ with a DFE of 0.65 eV, whereas the DFE of $\mathrm{V_{Sn}}^{2-}$ is 0.19 eV higher. Although this particular difference may be the result of the particular ordered structure we have chosen to represent $\mathrm{PEA_2Pb_{0.5}Sn_{0.5}I_4}$, we would argue that DFEs of metal vacancies in this compound are larger than in pure PEA$_2$SnI$_4$, provided Sn and Pb metal atoms are well mixed on an atomic scale. That this is the case is corroborated by optical measurements \cite{Fang2022}.
The DFEs of other defects, such as $\mathrm{V_{PEA}}^-$, are comparable to those in pure PEA$_2$PbI$_4$, or even higher, see \autoref{fig: DFEs}(f). In summary, mixing Pb and Sn in the 2D perovskite significantly suppresses the formation of defects compared to the pure Sn-based perovskite, and maintains the defect tolerance of the pure Pb-based perovskite.

\begin{figure}[h!]
    \includegraphics[width=0.8\textwidth]{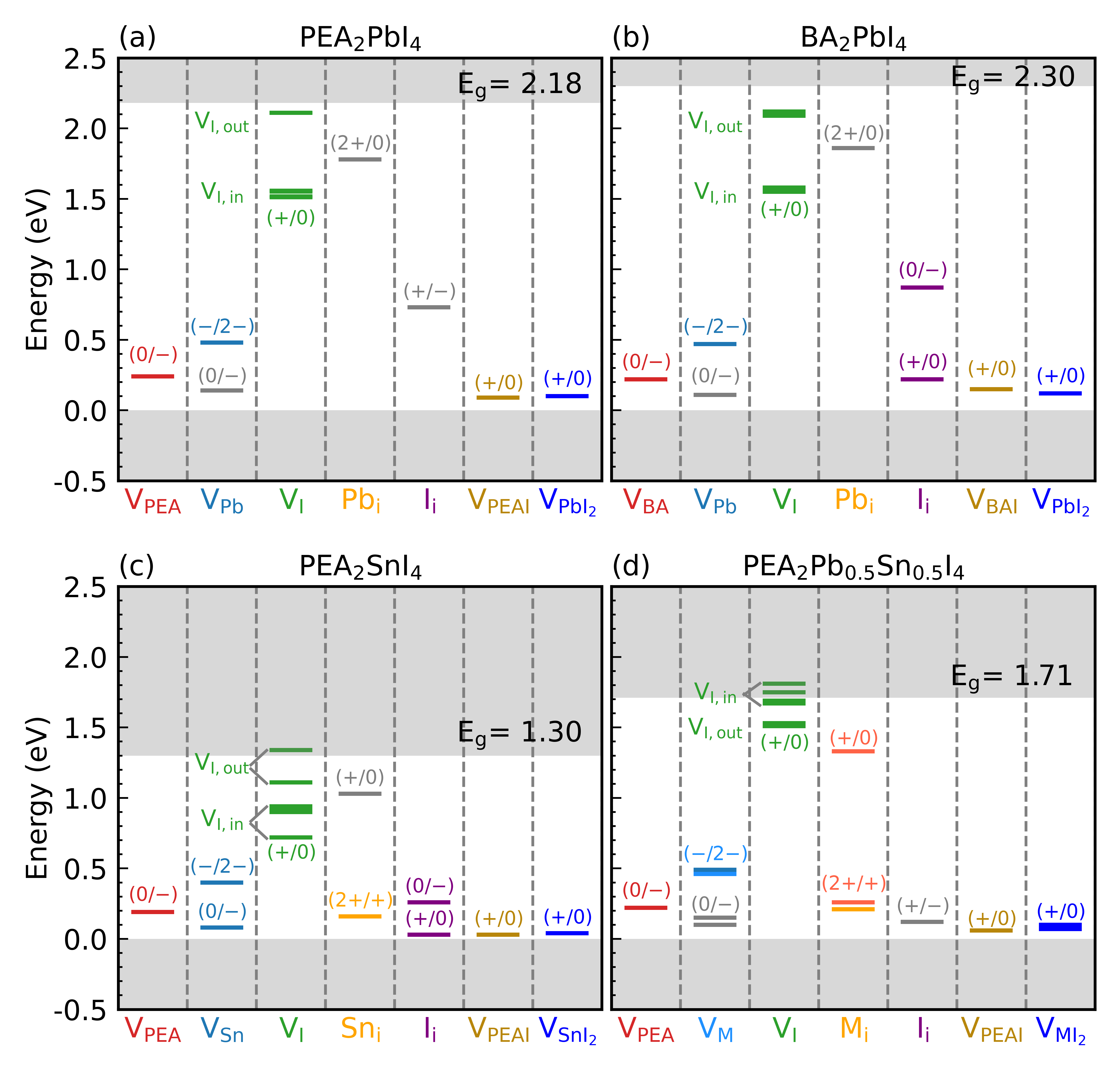}
    \caption{Charge state transition levels (CSTLs) of defects in PEA$_2$PbI$_4$ (a), BA$_2$PbI$_4$ (b), PEA$_2$SnI$_4$ (c), and PEA$_2$Pb$_{0.5}$Sn$_{0.5}$I$_4$ (d); the most important ones are indicated by colored lines, representing a change of a single unit $\pm e$ starting from one stable charge state of a defect; the bottom and top gray areas represent the valence and conduction bands (calculated with SCAN+rVV10), aligned at the VBM. In (d), M represents metal cations, Pb and Sn.}
    \label{fig: CSTLs}
\end{figure}

Although in equilibrium defect concentrations in the 2D perovskites considered here, with the exception of PEA$_2$SnI$_4$, are predicted to be small, materials are often grown under highly non-equilibrium conditions, during which a substantial amount of defects can form \cite{Liu2020_defect_density_2D_exp, Xue2022_compund_defects}. Likewise, under solar cell operating conditions, the (quasi) Fermi levels are very different from the intrinsic Fermi level, which may stimulate the formation of certain defects, \autoref{fig: DFEs}. These two points motivate investigating whether defects lead to electronic levels inside the band gap of a 2D material that can be harmful to its electronic operation.

These (so-called) charge state transition levels (CSTLs) are shown in \autoref{fig: CSTLs} for each point defect and compound vacancy in each of the 2D perovskites studied here. Generally, CSTLs are considered to be deep levels when their energy distance from the band edges is much larger than the thermal energy $k_B T$ (0.026 eV at the room temperature) \cite{Freysoldt2014}. Deep levels can trap charge carriers and cause significant (nonradiative) recombination, thereby reducing solar cell or light-emitting diode efficiencies. What can be observed in \autoref{fig: CSTLs} is that most point defects in 2D perovskites lead to deep levels. In fact, only the compound vacancies, $\mathrm{V_{PEAI}}$ and $\mathrm{V_{PbI_2}}$, give shallow (acceptor) levels. This is in remarkable contrast to what is found for 3D perovskites, where most, point or compound, defects produce shallow levels \cite{Xue2022perovskites, Xue2022_compund_defects}. This immediately provides a possible explanation of why the efficiency of 2D halide perovskite solar cells is generally smaller than those based on 3D perovskites.

For 3D perovskites it is argued that defects producing mainly shallow levels stems from the fact that defect states in these materials either have valence band or conduction band character, depending on the type of defect, and that the disruption in the chemical bonding pattern caused by the defect, is not so large as to move the defect levels in energy far from the band edges. \autoref{fig: CSTLs} shows that the CSTLs in 2D perovskites are still either in the top half or in the bottom half of the band gap, depending on the type of defect, but apparently the disruption caused by the defects is now sufficiently large to move defect levels to well inside the band gap. Part of these changes in going from 3D to 2D perovskites can also be explained by a decrease of the dielectric constant, which is also expected to lead to deeper levels.

As an example, consider the organic cation vacancies in 3D perovskites such as MAPbI$_3$, FAPbI$_3$ or MASnI$_3$, where $\mathrm{V_{MA}}^-$ and $\mathrm{V_{FA}}^-$ only give a very shallow acceptor level just above the VBM  \cite{Xue2022perovskites}. In all 2D compounds studies here, $\mathrm{V_{PEA}}^-$ or $\mathrm{V_{BA}}^-$ gives an acceptor level $\gtrsim 0.2$ eV above the VBM that can act as a trap state, see the first columns in \autoref{fig: CSTLs}(a)-(d). Likewise, the Pb and Sn vacancies, $\mathrm{V_{Pb}}^{2-}$, $\mathrm{V_{Sn}}^{2-}$, which only give very shallow acceptor levels in 3D perovskites \cite{Xue2022perovskites}, give a couple of deep trap levels in 2D perovskites, as can be observed from the second columns in \autoref{fig: CSTLs}(a)-(d).

An interesting case is the iodine vacancy $\mathrm{V_I}$, which in 3D perovskites typically only gives shallow acceptor levels. In 2D perovskites, $\mathrm{V_I}$ generates different levels according to the position of the vacancy, whether above/below or in a PbI$_2$ plane, see \autoref{fig: structures}(c,d). Vacancies associated with the in-plane positions generate deep levels that are $\sim 0.65$ eV below the CBM, whereas iodine vacancies located outside PbI$_2$ planes give donor levels that are much closer to the conduction band edge, which is correlated with the larger extent of bonding disruption on the 2D network induced by the former. In experiments, a broad emission of light with frequencies corresponding to the upper half of the band gap is frequently found in photoluminescence spectra of 2D perovskites, where the peak of the emission spectrum is approximately at $0.6$ eV below the CBM \cite{Kahmann2020, Yin2020}. We suggest that this emission may be associated with defect states created by iodine vacancies. The only other defect that gives a level in the upper half of the band gap, the Pb interstitial $\mathrm{Pb_i}$, is not likely to occur in appreciable quantities, \autoref{fig: DFEs}. Moreover, the Sn interstitial, $\mathrm{Sn_i}$, only gives a level close to the VBM, and the broad emission of the type discussed above is also observed in 2D Sn-based perovskites.

To conclude, we employ first-principles calculations to study the defect formation energies and charge state transition levels of intrinsic defects in Ruddlesden-Popper hybrid iodide 2D perovskites. We find that the equilibrium concentrations of point defects in the 2D perovskites PEA$_2$PbI$_4$,  BA$_2$PbI$_4$, and PEA$_2$SnI$_4$ are much lower than in comparable 3D perovskites, indicating improved material stability of 2D perovskites. The stability of 2D Sn iodide perovskites can be further enhanced by alloying with Pb. Moreover, unlike the prominence of interstitials in 3D perovskites, 2D perovskites are dominated by vacancies. The difficulty in forming defects in 2D perovskites is attributed to two factors. One is that the bonding disruptions by defects are more detrimental in 2D than in 3D networks. Another is that the dielectric constants are smaller for 2D perovskites. These factors also cause the formation of deep defect levels in the band gap of 2D perovskites. Consequently, should point defects emerge in sizable concentrations, then those defects can hamper the optoelectronic performance of the 2D perovskites. Finally, we suggest that the trap levels of iodine vacancies are responsible for the broad sub-bandgap emission in 2D perovskites observed in experiments.

\begin{acknowledgement}
H.X. acknowledges the funding from the China Scholarship Council (CSC) (No. 201806420038). Z.C. acknowledges funding from the Eindhoven University of Technology. S.T. acknowledges funding by the Computational Sciences for Energy Research (CSER) tenure track program of Shell and NWO (Project number 15CST04-2) and the NWO START-UP grant from the Netherlands.
\end{acknowledgement}

\begin{suppinfo}

Detailed computational approaches, including DFT calculations, structures, defect formation energy and charge state transition level; structures of mixed Pb-Sn perovskite.

\end{suppinfo}

\bibliography{Manuscript_2D_perovskite}

\end{document}


\tableofcontents

\clearpage

\section{Computational Approaches}

\subsection{DFT calculations} \label{sec: DFT calculations}
Density functional theory (DFT) calculations are performed with the Vienna \textit{Ab-Initio} Simulation Package (VASP) \cite{Kresse1993, Kresse1996, Kresse1996a}, employing the SCAN+rVV10 \cite{Peng2016} functional for electronic calculations and geometry optimization. This functional combines the strongly constrained and appropriately normed (SCAN) \cite{Sun2015} meta-generalized gradient approximation (meta-GGA) functional with the long-range van der Waals interactions from the revised Vydrova-van Voorhis non-local correlation functional (rVV10) \cite{Sabatini2013}. It has emerged as a reliable functional for calculating defect properties of metal halide perovskites in our previous work, Ref. \citenum{Xue2021functionals}.

The spin-orbit coupling (SOC) is omitted, as it has little effect on the formation energies of defects \cite{Xue2021functionals}. Our calculations use a plane wave kinetic energy cutoff of 450 eV, with a maximum cutoff energy of the plane-wave-basis set for all elements being 280 eV (see \autoref{table: cutoff convergence} for the convergence test), and a $\Gamma$-point only \textbf{k}-point mesh. The energy and force convergence criteria are set to 10$^{-4}$ eV and 0.02 eV/\AA, respectively. Spin-polarization is included in all calculations.

\begin{table}[h]
    \centering
    \caption{Convergence test of the cutoff energy of the plane-wave-basis set using the PEA$_2$PbI$_4$ unitcell. For each cutoff energy, the lattice volume and ionic positions are fully relaxed.}
    \label{table: cutoff convergence}
    \begin{tabular}{m{3.5cm}m{1.5cm}m{1.5cm}m{1.5cm}}
        \hline
        \multirow{2}{*}{Cutoff energy (eV)} & \multicolumn{3}{l}{Lattice constant (\AA)} \\
         & a & b & c\\
        \hline
        375 & 8.58 & 8.58 & 31.72 \\
        400 & 8.60 & 8.60 & 31.79 \\
        425 & 8.61 & 8.61 & 31.84 \\
        450 & 8.63 & 8.63 & 31.90 \\
        475 & 8.63 & 8.63 & 31.90 \\
        500 & 8.63 & 8.63 & 31.90 \\
        \hline
    \end{tabular}
\end{table}

\subsection{Structures} \label{sec: structures}

Structures of PEA$_2$PbI$_4$, BA$_2$PbI$_4$ and PEA$_2$SnI$_4$ are taken from the experimentally determined lattices from Refs. \citenum{Du2017, Billing2007, Gao2019}, respectively. PEA$_2$Sn$_{0.5}$Pb$_{0.5}$I$_4$ is constructed by substituting Pb with Sn in PEA$_2$PbI$_4$, see \autoref{sec: mixed Pb-Sn} for the detailed discussion about the substitution strategy. Then the pristine structures are reoptimized using the SCAN+rVV10 functional, including reoptimizing the volume of the unit cell. 

Defective structures are then created starting from 2$\times$2$\times$1 supercells, which contain 16 formula units per supercell, with 752 and 624 atoms for PEA- and BA-based perovskites, respectively. An interstitial is created by adding to the supercell a cation or an anion in a specific charge state, and then optimize the atomic positions within the supercell. Likewise, a vacancy is created by removing from the supercell a cation or an anion. 

\subsection{Defect formation energy} \label{sec: DFE}
The defect formation energy $\Delta H_f$ is calculated from the expression \cite{Walle2004}
\begin{equation}\label{eq: DFE}
        \Delta H_{f}(D^q) = E_\mathrm{tot}(D^q)-E\mathrm{_{tot}(bulk)} - \sum_{k} n_{i} \  \mu_{i} + q(E_{F}+ E_\mathrm{VBM}+ \Delta V),
\end{equation}
where $E_\mathrm{tot}{D^q}$ and $E\mathrm{_{tot}(bulk)}$ are the DFT total energies of the defective and pristine supercells, respectively, and $n_i$ and $\mu_i$ are the number of atoms and chemical potential of atomic species $i$ added to ($n_i > 0$) or removed from ($n_i<0$) the pristine supercell in order to create the defect. 

\begin{figure}[h]
    \includegraphics[width=0.85\textwidth]{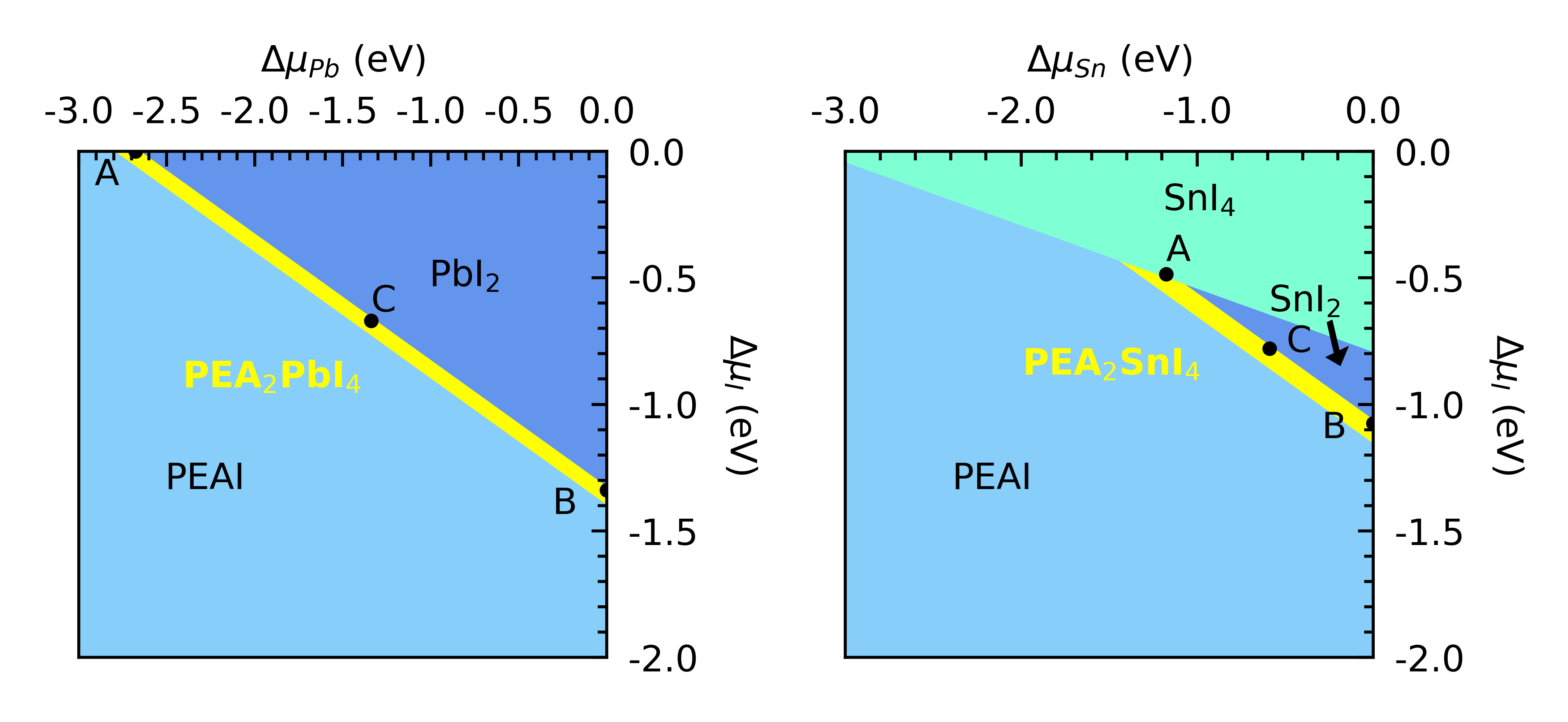}
    \caption{Calculated stability diagrams of PEA$_2$PbI$_4$ and PEA$_2$SnI$_4$. $\mu_\mathrm{I} = \mu_\mathrm{I_2,molecule}/2$ corresponds to  $\Delta \mu_\mathrm{I} = 0$ in the figures, and $\mu_\mathrm{M} = \mu_\mathrm{M,bulk}$ defines $\Delta \mu_\mathrm{M} = 0$ for M = Pb, Sn. The points A and B define iodine-rich and iodine-poor conditions, respectively. Iodine-medium conditions (point C) are defined as halfway between points A and B.}
    \label{fig: chemical potential}
\end{figure}

Creating a charge $q$ requires taking electrons from or adding them to a reservoir at a fixed Fermi level. The latter  is calculated as $E_{F}+ E_\mathrm{VBM}$, with $0\leq E_F \leq E_g$, the band gap, and $E_\mathrm{VBM}$ the energy of the valence band maximum. As it is difficult to determine the latter from a calculation on a defective cell, one establishes $E_\mathrm{VBM}$ in the pristine cell, shifted by $\Delta V$, which is calculated by lining up the core level on an atom in the pristine and the neutral defective cell that is far from the defect \cite{Walle2004, Komsa2012}. We neglect the vibrational contributions to the DFEs, and the effect of thermal expansion on the DFEs, as these are typically small in the present compounds \cite{Xue2021functionals, Wiktor2017}.

Taking PEA$_2$PbI$_4$ for example, the chemical potentials $\mu_i$ of atomic species $i$ are calculated by assuming that the perovskite is stable, so using 2$\mu _\mathrm{PEA} + \mu _\mathrm{Pb} + 4 \mu _\mathrm{X} = \mu_\mathrm{PEA_2PbI_4}$ as a constraint, where for $\mu_\mathrm{PEA_2PbI_4}$ we use the DFT total energy per formula unit of the PEA$_2$PbI$_4$ perovskite. Furthermore, we assume that the perovskite is in equilibrium with the PbI$_2$ phase, so $\mu_\mathrm{Pb} + 2 \mu_\mathrm{I} = \mu_\mathrm{PbI_2}$, with $\mu_\mathrm{PbI_2}$ the DFT total energy per formula unit of $\mathrm{PbI_2}$. All $\mu_i$ can now be expressed in terms of a single parameter, $\mu_\mathrm{I}$, which is constrained by $\mu_\mathrm{PbI_2} - \mu_\mathrm{Pb,bulk} \leq 2 \mu_\mathrm{I} \leq \mu_\mathrm{I_2,molecule}$. The outer bounds define I-poor (or Pb-rich) or I-rich conditions, respectively, with $\mu_\mathrm{Pb,bulk}$ and $\mu_\mathrm{I_2,molecule}$ the DFT total energies of bulk Pb metal, and an I$_2$ molecule. I-poor and I-rich conditions are indicated by points B and A, respectively, in \autoref{fig: chemical potential}(a). The I-medium condition (point C) is defined as the halfway between points A and B, which is the focus of this work. The stability diagram of BA$_2$PbI$_4$ is similar to PEA$_2$PbI$_4$. For PEA$_2$SnI$_4$, the allowed interval for $\mu_\mathrm{X}$ has to be narrowed down to prevent the formation of other phases, such as SnI$_4$ \cite{Shi2017, Meggiolaro2020a, Xue2022perovskites}, see \autoref{fig: chemical potential}(b). For PEA$_2$Sn$_{0.5}$Pb$_{0.5}$I$_4$, we take the point C in the stability diagram of PEA$_2$SnI$_4$ to get the $\mu_\mathrm{I}$ and $\mu_\mathrm{Sn}$. With the $\mu_\mathrm{I}$, the $\mu_\mathrm{Pb}$ is then determined in \autoref{fig: chemical potential}(a).

The intrinsic Fermi level can be determined by the charge neutrality condition, which expresses the fact that, if no charges are injected in a material, it has to be charge neutral
\begin{equation} \label{eq:charge_neutrality}
    p - n + \sum_{D^q} q \; c(D^q) = 0,
\end{equation}
where $p$ and $n$ are the intrinsic charge densities of holes and electrons of the semiconductor material, $c(D^q)$ is the concentration of defect $D^q$, and the sum is over all types of charged defects. The concentrations can be calculated from Boltzmann statistics
\begin{equation} \label{eq: defect concentration}
    c(D^q)=c_0(D^q)  \exp \left[- \frac{\Delta H_f (D^q)}{k_BT} \right],
\end{equation}
where $c_0(D^q)$ is the density of possible sites for the defect (defined by the number of possible sites for the defect $D^q$ in the unit volume), $T$ is the temperature, $k_B$ is the Boltzmann constant, and $\Delta H_f (D^q)$ follows from \autoref{eq: DFE}. Obviously, $p$, $n$, and $c(D^q)$ are functions of $E_F$, so the charge neutrality condition, \autoref{eq:charge_neutrality}, serves to determine the intrinsic position of the Fermi level $E_F^{(i)}$. 

Since the $E_F^{(i)}$ is relatively close to the valence band maximum, the intrinsic hole density $p$ is large and important for maintaining the charge neutrality together with the charged defects, while the intrinsic electron density $n$ is negligibly small. The hole concentration $p$ can be calculated from the density of states near the VBM:
\begin{equation} \label{eq: hole concentration}
    p = N_V^{2D}\frac{1}{l_z} \exp \left[- \frac{E_F}{k_BT} \right], \, \mathrm{where} \, N_V^{2D} = 2 \frac{2 \pi m_h^* k_B T}{h^2}.
\end{equation}
In \autoref{eq: hole concentration}, $N_V^{2D}$ is the 2D effective density of states of the valence band edge, which can be modeled assuming parabolic bands, using the effective hole masses $m_h^*$ at the VBM. The $m_h^*$ for PEA$_2$PbI$_4$, BA$_2$PbI$_4$ and PEA$_2$SnI$_4$, taken from Ref. \citenum{Dyksik2020}, are 0.25$m_0$, 0.39$m_0$, and 0.15$m_0$, respectively, where $m_0$ is the mass of the electron. For PEA$_2$Sn$_{0.5}$Pb$_{0.5}$I$_4$ the same value as PEA$_2$SnI$_4$ is used. The $N_v^{2D}$ is divided further by the thickness of one layer of the 2D perovskite, $l_z$, defining the 3D effective density of states.

\subsection{Charge state transition level} \label{sec: CSTL}

Under operating conditions, charges are injected in the material, shifting the positions of the (quasi) Fermi levels for electrons and holes. The charge state transition level (CSTL) $\varepsilon(q/q')$ is defined as the Fermi level position where the charge states $q$ and $q'$ of the same type of defect have equal formation energy, $\Delta H_{f}(D^q)=\Delta H_{f}(D^{q'})$. As the DFEs have a simple linear dependence on $E_F$, \autoref{eq: DFE}, this condition can be expressed as
\begin{equation} \label{eq:CSTL}
    \varepsilon(q/q')= \frac{\Delta H_{f}(D^q,E_{F}=0)-\Delta H_{f}(D^{q'},E_{F}=0)}{q'-q},
\end{equation}
where $\Delta H_{f}(D^q,E_{F}=0)$ is the DFE calculated at $E_F = 0$. 

\newpage

\section{Mixed Pb-Sn perovskite} \label{sec: mixed Pb-Sn}

\begin{figure}[h]
    \includegraphics[width=0.5\textwidth]{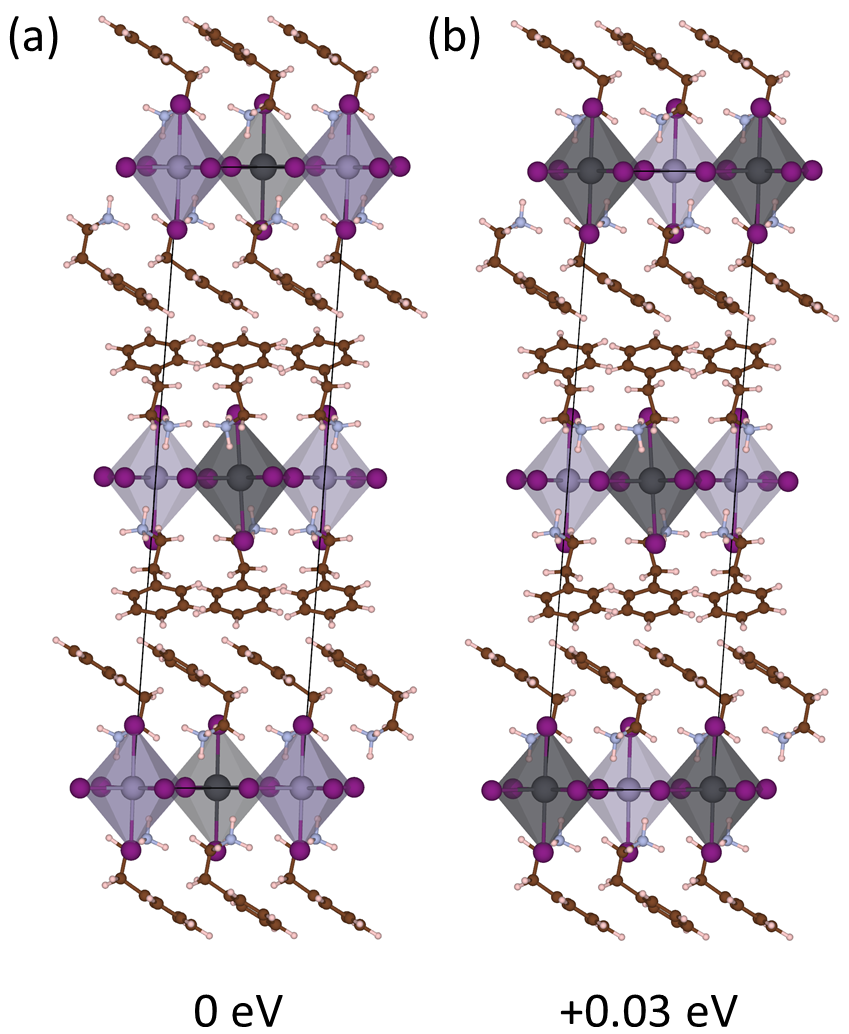}
    \caption{Optimized structures of two possible configurations of $\mathrm{PEA_2Pb_{0.5}Sn_{0.5}I_4}$ with well mixed Pb and Sn.}
    \label{fig: mix Pb-Sn}
\end{figure}

The main interaction of ions in 2D perovskites is within the inorganic layer. Therefore, Pb and Sn are mixed in the same layer. Whereas the stacking pattern of adjacent layers has two possibilities, as shown in \autoref{fig: mix Pb-Sn}, the second configuration is 0.03 eV per unitcell less stable than the first one. Therefore, we choose the one in \autoref{fig: mix Pb-Sn}(a) as the model for studying defects. Another possible configuration can be one pure Sn layer stacked with one pure Pb layer, but apparently the local interaction is expected to be similar to the pure Sn- or Pb-based perovskites.

\newpage

\bibliography{SI}